\documentclass{svmult}
\usepackage{amsmath,amsfonts,amssymb,amscd,colordvi}
\usepackage{graphicx}
\usepackage{enumerate}
\usepackage{latexsym}
\usepackage[usenames]{color}
\renewcommand{\Re}{\mathop{\rm Re}\nolimits}
\renewcommand{\Im}{\mathop{\rm Im}\nolimits}
\newcommand{\p}{\partial}

\newcommand{\e}{\varepsilon}

\newcommand{\vp}{\varphi}

\newcommand{\gi}{\delta}

\newcommand{\lla}{\gamma}

\newcommand{\bk}{{\mathbf k}}

\newcommand{\bx}{{\mathbf x}}
\newcommand{\be}{\begin{equation}}
\newcommand{\ee}{\end{equation}}

\newcommand{\bb}{\mbox{\boldmath$\beta$}}
\newcommand{\R}{{\mathbb R}}
\newcommand{\C}{{\mathbb C}}

\newcommand{\Z}{{\mathbb Z}}

\newcommand{\T}{{\mathbb T}}

\newcommand{\ai}{a}

\newcommand{\cM}{{\cal M}}

\newcommand{\cT}{{\cal T}}

\newcommand{\const}{\mathop{\rm const}\nolimits}

\newcommand{\sgn}{\mathop{\rm sgn}\nolimits}
\newcommand{\Arg}{\mathop{\rm Arg}\nolimits}

\renewcommand{\vec}[1]{{\mathbf{#1}}}

\def\12{\tfrac12}

\def\lan{\langle}
\def\ran{\rangle}

\def\eps{\varepsilon}
\def\p{\partial}

\def\e{\varepsilon}

\def\o{\omega}

\def\be{\begin{equation}}\def\ee{\end{equation}}
\def\bea{\begin{eqnarray}}\def\eea{\end{eqnarray}}

\begin{document}

\author{Sergei Kuksin\inst{1}
\and
Alberto Maiocchi\inst{2}}
\institute{Paris\and Milan}
\authorrunning{S. Kuksin, A. Maiocchi}

\title{The effective equation method}
%\date{}
%%\date{(preliminary version)}
\maketitle
\label{chp:kuksin}

\abstract{In this chapter we present a general method of constructing the 
effective equation which  describes the behaviour of small-amplitude 
solutions for a nonlinear PDE in finite volume, provided that the linear part of 
the equation is a hamiltonian system with a pure imaginary discrete spectrum.
The effective equation is obtained by retaining only the resonant terms of the nonlinearity
(which may be  hamiltonian, or may be not); the assertion that it describes the 
limiting behaviour of small-amplitude solutions is a rigorous mathematical 
theorem. In particular, the method applies to the 
three-- and four--wave systems. We demonstrate that different possible types of 
energy transport are covered by this method, depending on whether the set of resonances
splits into finite clusters (this happens, e.g. in case of the Charney-Hasegawa-Mima equation),
or is connected (this happens, e.g. in the case of the NLS equation if the space-dimension is
at least two). For equations of the first type the energy transition to high frequencies does not hold, while for equations of the second type it may take place. In the case of the NLS equation we use next 
some heuristic 
approximation from the arsenal of wave turbulence to show that under the iterated limit
``the volume goes to infinity", taken after the limit ``the amplitude of oscillations goes to zero", the 
energy spectrum of solutions for the effective equation is described by a Zakharov-type kinetic 
equation. Evoking the Zakharov ansatz we show that stationary in time and homogeneous in
space solutions for the latter equation have a power law form.  Our method applies to  various weakly
nonlinear wave systems, appearing in plasma, meteorology and oceanology.}

\section{Introduction}
It is well known that solutions of 
linear evolution PDEs in finite volume 
 are superpositions of normal modes of oscillations (in most cases of interest
these are the Fourier modes). When a nonlinearity is added as a perturbation, 
different modes start to  interact and the
solutions of the equation can be approximated by 
suitable power series expansions, provided that the nonlinearity is
sufficiently small (or, in other words, the PDE is \emph{weakly}
nonlinear). In such cases, the equation can be written as 
 \be \label{ini}
 u_t+
 L(u)=\e N(u)\,,
 \ee
 where $L$ is the linear operator, $N$ denotes the nonlinearity and $\e$
 is a small parameter, $0<\e \ll 1$. The equation may contain a stochastic 
 force, and in that case it reads
 \be \label{1.1}
 u_t+
 L(u)=\e N(u) +\sqrt\e\,\lan\text{random force}\ran
 \ee
 (the scaling of the random force by the factor $\sqrt\e$ is the most natural, see below). 
 We will show that the limiting, as $\e\to0$,  exchange
 of energy between the modes may be  described by replacing the original system with a 
 suitable 
 \emph{effective equation}. This result may be regarded as a PDE-version of the Bogolyubov 
 averaging principle (see  \cite{BogMitr}) which implies a similar property for distribution
 of energy between the oscillating modes for small-amplitude oscillations in 
 finite-dimensional nonlinear systems. 

The mentioned above convergence that holds as $\e\to0$ and various properties of the 
corresponding effective equations have been rigorously established
 (see \cite{K10, Kuk11, KM13,KM13KZ, bplane, wnpdes} and the discussion in \cite{wnpdes}).  
 The treatments of the 
deterministic and stochastic equations  are similar,  but the results, obtained in the stochastic 
case, are significantly stronger: while the deterministic effective equation 
 controls the dynamics only on time intervals of order 
 $\eps^{-1}$, in the presence of stochastic forcing the corresponding effective 
  equation also 
 approximates  the stationary measure for  the original
equation, thus controlling  the asymptotical in time 
 behaviour of   solutions when $\e\ll1$.  Moreover, in
the absence of forcing we only get information concerning  the exchange of
energy between  the modes, whereas in the stochastically 
 forced case, the stationary
measure for the effective equation controls both the energies and the
phases of the normal modes of solutions.

Below we explain  how to construct the effective equations for eq. \eqref{ini}
and eq.~\eqref{1.1} from   the resonant terms of the nonlinearities. We will 
discuss two   examples:  the nonlinear Schr\"odinger
  equation and the Charney-Hasegawa-Mima equation on the $\beta$
  plane. These two equations 
   display completely different types of  energy exchange between
  modes, and we will explain why this happens.

\section{How to construct the effective equation}\label{sec:ee}
We  consider  hamiltonian PDEs, whose linear parts
have  imaginary pure point spectra  and are diagonal in Fourier modes.
Written in terms of the complex Fourier coefficients  $v=\{v_\bk\}$
(also called waves), the equations which we study read
%The most general form we will consider  is that of a system
%of equations for the Fourier coefficients $v=\{v_\bk\}$ of the kind
\begin{equation}\label{eq:or*}
\frac{d}{dt} v_\bk=i\omega_\bk v_\bk +P_\bk(v)\ ,\qquad \bk\in \Z^d\ .
\end{equation}
Here  $\omega_\bk$ are real numbers
 and $P(v)= (P_\bk(v), \bk\in\Z^d)$ is a polynomial
nonlinearity in $v$ of certain order $q$,  of the form 
$$
P_\bk(v)=\sum_{p\le q}\sum_{\bk_1\ldots \bk_p\bk_{p+1}\ldots \bk_q}
c_{\bk_1\ldots \bk_q\bk} v_{\bk_1} \cdots 
v_{\bk_p} v^*_{\bk_p+1}\cdots  v^*_{\bk_q} \delta^{1\ldots
  p}_{p+1\ldots q\,\bk}\ ,
$$
where   $c_{\bk_1\ldots \bk_q\bk}$ are some complex
coefficients, $v^*$ is the complex conjugate of $v$ and
\be \label{Kronecker}
\delta^{1\ldots  p}_{p+1\ldots q\bk}=\left\{\begin{array}{cc}
1& \mbox{if } \bk_1+\ldots+\bk_p= \bk_{p+1}+\ldots+\bk_q+\bk\\
0& \mbox{else}
\end{array}
\right.\ .
\ee
We always assume that ``the nonlinearity does not pump energy in the system":
\be\label{dissipate}
\text{Re}\, \sum_\bk P_\bk(v)\bar v_\bk  \le0
\ee
(in most case of interest the l.h.s. vanishes).

The quantities $I_\bk:=|v_\bk|^2/2$, $E_\bk=\o_\bk I_\bk$ and
$\varphi_\bk=\Arg(v_\bk)$ are called, respectively, the 
wave action, wave energy and wave
phase.  The relation between $\omega$ and $\bk$, i.e. the function 
$\bk\to\omega_\bk$,
is
 called the {\it  dispersion relation}, or {\it dispersion function}.

The  {\it weakly nonlinear regime}  corresponds to solutions of small
amplitude $\eps$. We will  study it  in the presence of damping and, possibly, a random force, whose
magnitude is controlled by another parameter, call it $\nu$. So,
instead of \eqref{eq:or*}, we will consider 
\begin{equation}\label{eq:or}
\frac{d}{dt}v_\bk=i\omega_\bk v_\bk +\eps^q P_\bk(v)-\nu\gamma_\bk v_\bk
+\mu\sqrt{\nu}\,b_\bk \dot \bb_\bk\ ,\qquad \bk\in \Z^d\ ,
\end{equation}
where $\gamma_\bk\ge\gamma_*  >0$ controls the damping term, $b_\bk>0$ controls the forcing
and the parameter $\mu\in\{0,1\}$ is introduced to consider at the same time the 
forced and non-forced cases. The $\dot\bb_\bk$'s are complex white noises, independent  from each
other.\footnote{ That is, $\dot\bb_\bk=(d/dt)\bb_\bk$, \  
$\bb_\bk=\bb_\bk^+ + i \bb\bk^-$, where $\bb_\bk^\pm$ are standard independent 
real Wiener processes.}
The factors $\nu$ and $\sqrt \nu$ in front of the damping and the
dissipation are so chosen that, in the limit of $\nu\to 0$, the
solutions stay of order one, uniformly in $t>0$. 

Note that, while $\eps$ controls the size of the solutions, $1/\nu$
is the time-scale on which the forcing acts significantly, as it is
the time needed for the standard deviations of the processes $\sqrt \nu
\bb_\bk$ to become of order one. 
If $\mu=0$ and the system \eqref{eq:or}
is deterministic, still its time-scale is $1/\nu$ since, as we explain below,
 its solutions with initial data of order one 
 stay of order one  for $t \lesssim\nu^{-1}$, 
 while for much bigger values 
of time they  are very small   since in view of \eqref{dissipate} 
their $\ell_2$-norms   satisfy 
%of \eqref{eq:or}, where ${\mu=0}$,
$$
|v(t)|_{\ell_2}^2 \le |v(0)|_{\ell_2}^2 e^{-t\nu\gamma_*}\,.
$$

We will consider the regime
\begin{equation}\label{regime}
\nu=\eps^q\ 
\end{equation}
(where $q$ is the degree of $P$), 
and study  solutions of the equation with given   initial conditions
 on the time-scale
$1/\nu$, examining them under the limit $\nu\to 0$.
Passing to the slow time $\tau=\nu t$ (so that time 
$t\sim 1/\nu$ corresponds to $\tau$ of order 1),
eq. \eqref{eq:or} becomes
\begin{equation}\label{eq:or2}
\dot v_\bk=i\nu^{-1}\omega_\bk v_\bk +P_\bk(v)-\gamma_\bk v_\bk
+\mu b_\bk \dot \bb_\bk\ , \qquad\bk\in \Z^d\ ,
\end{equation}
where the upper dot  stands for $\tfrac{d}{d\tau}$.

We claim that, in the limit when  $\nu$ (or, equivalently, $\eps$)
goes to zero, the distribution of the energies $E_\bk$ 
 on times $\tau$ of order one is described
by an \emph{effective equation} whose nonlinearity is 
constituted by  resonant terms of
the nonlinearity (see \eqref{eq:ee} below).
\medskip

It is easier to understand the 
 role of  resonances and the form of the effective equation
  by passing to the interaction representation ( cf. \cite{BogMitr, ZLF92, Naz}), 
i.e., by performing the time-dependent change of variables from $v_\bk$
to
\be\label{ak}
\ai_\bk=e^{-i\nu^{-1} \omega_\bk \tau} v_\bk\ ,
\ee
which transforms \eqref{eq:or2} to
\begin{equation}\label{eq:int}
\dot \ai_\bk=\mathbf R_\bk(\ai,\nu^{-1}\tau)-\gamma_\bk \ai_\bk
+\mu b_\bk e^{-i\nu^{-1} \omega_\bk \tau} \dot \bb_\bk\ , \qquad\bk\in \Z^d\ ,
\end{equation}
where $\mathbf R_\bk$ denotes the nonlinearity, written 
 in the $\ai$-variables. That is
\begin{equation*}
\begin{split}
\mathbf R_\bk(\ai,\nu^{-1}\tau)=& \sum_{p\le q}\sum_{\bk_1\ldots
  \bk_p\bk_{p+1}\ldots \bk_q} c_{\bk_1\ldots \bk_q\bk} v_{\bk_1}
\cdots  v_{\bk_p}  v^*_{\bk_p+1}\cdots  v^*_{\bk_q} \delta^{1\ldots
  p}_{p+1\ldots q\,\bk}\\
& \times \exp\left(i\nu^{-1}\tau\left(\omega_{\bk_1}+\ldots+\omega_{\bk_p}-
  \omega_{\bk_{p+1}} -\ldots
  -\omega_{\bk_q}-\omega_{\bk}\right)\right)\ .
\end{split}
\end{equation*}
Noting that  the collection of the processes
$
\{e^{-\nu^{-1}\omega_\bk\tau}\dot\beta_\bk\}
$
is another set of standard independent complex white noises, we  re-write
eq.~\eqref{eq:int} as 
\begin{equation}\label{eq:int2}
\dot \ai_\bk=\mathbf R_\bk(\ai,\nu^{-1}\tau)-\gamma_\bk \ai_\bk
+\mu b_\bk \dot \bb_\bk\ , \qquad\bk\in \Z^d\ .
\end{equation}

In the sum defining $\mathbf R_\bk$,  the terms for which the {\it resonance
conditions}
\begin{equation}\label{resonance}
\left\{\begin{array}{c}
\omega_{\bk_1}+\ldots+\omega_{\bk_p}=  \omega_{\bk_{p+1}} +\ldots
+\omega_{\bk_q}+\omega_{\bk}\\
\bk_1+\cdots+\bk_p=\bk_{p+1}+\cdots \bk_q+\bk
\end{array}
\right.
\end{equation}
are satisfied  (called  the {\it resonant terms})  under the
limit $\nu\to 0$ behave  completely differently  from  others terms 
(called the {\it nonresonant terms}). Namely, the 
nonresonant terms oscillate faster
and faster, whereas the resonant terms do not. We will say
that a set of $\Z^d$-vectors $\{\bk_1,\ldots,\bk_q,\bk\}$ forms a
\emph{resonance} if relations \eqref{resonance} are satisfied, if 
$c_{\bk_1\ldots\bk_q\bk}\neq 0$, and the set 
$\{\bk_1,\ldots,\bk_p\}$ does not  equal the set 
$\{\bk_{p+1},\ldots,\bk_q,\bk\}$. The collection  of all resonances is called the 
\emph{resonant set}.

In the spirit of the finite-dimensional 
averaging, following the Bogolyubov averaging principle
 (see \cite{BogMitr}, the behaviour of solutions of \eqref{eq:int2}  under the limit
$\nu\to 0$ is obtained by replacing the 
nonlinearity $\mathbf R_\bk$ with its time average, i.e. with 
$$
\lim_{T\to \infty} \frac 1T\int_0^T \mathbf R_\bk (a,t) dt\ .
$$
Since for any real number $\lambda$ we have 
$$
\lim_{T\to \infty} \frac 1T\int_0^T e^{i\lambda t}  dt=
\left\{
\begin{array}{cc}
1& \mbox{if }\lambda =0\\
0& \mbox{if }\lambda\neq 0
\end{array}
\right.\ ,
$$
then only the resonant terms survive in the averaged nonlinearity. We
write their sum as 
\begin{equation*}
R_\bk(\ai)=\sum_{p\le q} \sum_{\bk_1\ldots
  \bk_p\bk_{p+1}\ldots \bk_q} c_{\bk_1\ldots \bk_q\bk} v_{\bk_1}
\cdots  v_{\bk_p}  v^*_{\bk_p+1}\cdots  v^*_{\bk_q} \delta^{1\ldots
  p}_{p+1\ldots q\,\bk}\delta(\omega^{1\ldots p}_{p+1\ldots q\,\bk})\ ,
\end{equation*}
where
$$
\delta(\omega^{1\ldots  p}_{p+1\ldots q\bk})=\left\{\begin{array}{cc}
1& \mbox{if } \omega_{\bk_1}+\ldots+\omega_{\bk_p}=
\omega_{\bk_{p+1}}+\ldots+\omega_{\bk_q}+\omega_\bk\\
0& \mbox{else}
\end{array}
\right.\ .
$$

This suggests to take for the  effective equation the following system:
\begin{equation}\label{eq:ee}
\dot {\tilde a}_\bk=R_\bk(\tilde a)-\gamma_\bk \tilde a_\bk
+\mu b_\bk \dot \bb_\bk\ , \qquad\bk\in \Z^d\ .
\end{equation}
Indeed, it is proved in \cite{wnpdes} (also see \cite{K10, Kuk11, KM13}) 
that, if the original equation
\eqref{eq:or} is well posed on time intervals $t\lesssim 1/\nu$, then 
eq.~\eqref{eq:ee} describes the limiting behaviour of the variables $\ai_\bk$
(and, as well, the distribution of energy  since $|v_\bk|=|\ai_\bk|$) in the
time-scale $t\sim 1/\nu$,  for any sufficiently regular initial data. This
holds both in the presence and in the absence of the random forcing (i.e., both
for $\mu= 0$ and $\mu=1$). Moreover, in the forced case we also control the 
limiting behaviour  of the stationary solutions for eq.~\eqref{eq:or}. So if  the
equation \eqref{eq:or} and the effective equation both are mixing, then we control the
behaviour of all solutions for \eqref{eq:or}  under the iterated limit
$\ 
\lim_{\e\to0}\, \lim_{t\to\infty}\,.
$ 
Remarkably, in this case the effective equation describes not only the limiting
behaviour of the  actions, but also that of the 
angles. I.e., it completely controls the limiting distribution of solutions. So
if $f(v)$ is a functional on the space of sequences $v=(v_\bk)$, satisfying 
some mild restriction on its growth as  the norm of $v$ goes to infinity, 
 %$|v|_{\ell_2}\to\infty$,
  and $v^\e(t)$ is any 
solution for \eqref{eq:or}, then 
$$
\lim_{\e\to0}\, \lim_{t\to\infty} {\mathbf  E} f(v^\e(t)) \to \int f(v) \, \mu(dv)\,,
$$
where $\mu$ is the unique stationary measure for the effective equation 
\eqref{eq:ee} and ${\mathbf  E}$ signifies the expectation. 
See in \cite{Kuk11, KM13,wnpdes}.

\section{ Structure of resonances}\label{sec:res}

We intend to use the effective equations as a tool to investigate the 
energy transport in different physically relevant PDEs. We will show
 that the limiting, as $\e\to0$, energy transport
   for any specific equation  depends on the structure of
the resonances (which, in turn, is determined by the form of the
dispersion function $\omega_\bk$).

Three possibilities can occur:

\noindent 
1)  The resonant set is empty. Then if the degree $q$ of the nonlinearity is even,
  the effective equation is linear. If  $q$ is odd,  the equation 
   may contain  nonlinear integrable  terms  of the form $f(I)v_\bk$. But 
 these terms  do not contribute to the
  dynamics of the wave actions. So in any case 
   different modes do not exchange energy,
  and no energy transport to high frequencies   occurs. 
  \smallskip
  
  Now assume that the resonant set is not empty. We say that integer 
  vectors $\bk_1, \bk_2 \in\Z^d$ are {\it equivalent} if there exist vectors 
  $\bk_3,\dots,\bk_q, \bk \in\Z^d$,  such that the relations 
   \eqref{resonance} hold. This equivalence divides $\Z^d$ to
   clusters, formed by 
   elements which can be joined by chains of equivalences (see 
   \cite{Kar,Kar94} for a discussion of the role of resonant clusters
   in weak turbulence). 
   
   The two remaining
   cases are:
  \smallskip
  
  \noindent 
2)  All resonances are connected, so the whole $\Z^d$ is a single
cluster.
%that no isolated clusters of  resonances are  contained in the resonant set. 
 In this case, in the   limit when the volume of the space-domain goes to infinity,
 under some additional  assumptions
    a new  type of kinetic  equation can be derived, the energy transport 
    takes place     and power law  stationary spectra, which   depend
    only on the form of the 
     dissipation, can be obtained.
  \smallskip
  
  \noindent 
3)  All resonances are divided to  non intersecting clusters. Now
 the energy transfer should be studied separately within each
  cluster. If sizes of    the  clusters are bounded, then no 
  energy transport to high frequencies  occurs. 
  \medskip 
  
  See \cite{Kuk11, hgdcds} for the case 1), \cite{KM13, KM13KZ} for the stochastic case~2)
  and \cite{hgjdde} for the deterministic case, 
   and see \cite{bplane} for the stochastic case~3).
  See \cite{wnpdes} for discussion and for  theorems, applicable in all three cases, deterministic 
  and stochastic. 
  
  Note that many examples of systems which fall to type 2) are given by equations
 \eqref{eq:or} with completely resonant spectra $\{\omega_\bk\}$, i.e. with spectra of the form
 $\omega_\bk=\omega_*\,\Omega_\bk$, where $\Omega_\bk$ are integers.  Averaging theorems
 for completely resonant deterministic  equations  \eqref{eq:or} with $\nu=0$ were discussed 
 in \cite{ FGH, GG12, hgjdde}; also see \cite{wnpdes}. 

Below we discuss examples  for the case 2) when  all
resonances are connected (Section~\ref{sec:NLS}), and for  the case~3)  when 
the resonances make  non intersecting finite  clusters (Section~\ref{sec:CHM}). For
more examples of systems of types 2) and 3) see \cite{Kar}. 

\subsection{The equations}
Our first example is the cubic NLS equation  on the
$d$-dimensional torus of size $L$ (see
\cite{KM13,KM13KZ}):
\begin{equation}\label{eq:ZL*}
\p_t u-i\Delta u= i\eps^2\gi\left|u\right|^2u+ 
\nu\langle\text{
dissipation}\rangle +\mu\sqrt\nu\langle \text{random forcing}\rangle\,, 
\quad u=u(t,\bx)\in \C\ ,
\end{equation}
for $\bx\in \T^d_L=\R^d/(2\pi L\Z^d)$, where the parameter $\gi=\gi(L)$
is introduced in order to control different scaling for solution as
the size $L$ of the torus varies.\footnote{More exactly, \eqref{eq:ZL*} is 
 the damped/driven cubic NLS equation.
See \cite{ZLF92, Naz} for the non-perturbed NLS equations.}  The dissipation is a linear
operator, diagonal in the  
exponential base of  functions on $\T^d_L$ 
\be\label{Four}
\{\phi_\bk(x)= e^{iL^{-1} \bk\cdot x},\ \bk\in\Z^d\}\,.
\ee
As before, by $v=\{v_\bk, \bk\in\Z^d\}$ we denote the the Fourier coefficients of $u(x)$:
$$
u(x) = \sum_{\bk\in\Z^d} v_\bk \phi_\bk(x)\,.
$$

If $d=1$, the resonance condition \eqref{resonance} 
takes the form
$$
k_1^2+k_2^2= k_3^2+k^2,\qquad k_1+k_2=k_3+k\,.
$$
All solutions for this system 
 are such that $k_1=k, k_2=k_3$, or $k_1=k_3, k_2=k$. So in this
case the resonant set is empty, and no energy cascade to high frequencies happens
when $\e^2=\nu\to0$.  This is well known.

Now consider a
 higher-dimensional NLS equation, %are examples  
 % of  wave systems in which all resonances are connected. To see this 
% let us 
  write it  in  the Fourier variables $\{v_\bk$, 
$\bk\in \Z^d\}$, and  pass  to the slow time $\tau=\nu t$. Then,
if the forcing and the dissipation are chosen in accordance with
the prescription of the previous section (cf. \eqref{eq:or}), the equation reads 
%equation reads (see sec.~\ref{sec:ee})
\begin{equation}\label{eq:NLS}
\dot v_\bk=-i\nu^{-1}\omega^N_\bk v_\bk
+i\gi\sum_{\bk_1,\bk_2,\bk_3\in \Z^d}
v_{\bk_1} v_{\bk_2}v^*_{\bk_3} \delta^{12}_{3\bk} -\gamma_\bk v_\bk
+\mu b_\bk \dot \bb_\bk\,.
\end{equation}
 Here
$-\gamma_\bk$ are the eigenvalues of the dissipation operator. The 
 eigenvalues of the operator $-\Delta$, call them
$\omega^N_\bk$, follow the dispersion relation
\begin{equation}\label{eq:dispNLS}
\omega^N_\bk=\left|\bk\right|^2/L^2\ ,\quad \bk\in \Z^d\ .
\end{equation}
Below we will see that if $d\ge2$, then the whole $\Z^d$ forms a single cluster,
so the equation fits the case~2).
\medskip

An interesting example of the case 3) of isolated clusters
is provided by the Charney--Hasegawa--Mima (CHM)
equation on the $\beta$ plane  (see \cite{bplane} and see \cite{ZLF92,Naz} for this 
equation with $\nu=0$), which we
write as
$$
(-\Delta+F) \p_t\psi-\eps J(\psi,\Delta \psi)-\p_x \psi=
\nu\langle\text{
dissipation}\rangle +\mu\sqrt\nu\langle \text{random forcing}\rangle\,, 
% \nu\langle diss.\rangle +\sqrt\nu\langle forc.\rangle\ , 
\quad
\psi=\psi(t,\bx)\in \R\ .
$$
Here the constant 
 $F\ge 0$ is called the Froude number and $J(\psi,\Delta \psi)$
denotes the Jacobian determinant of the vector $(\psi,\Delta \psi)$. The 
space-domain is a strip of horizontal size $\rho$ and
vertical size one, under double periodic boundary conditions, i.e.,
$$
\bx=(x,y)\in \T^2_{\rho,1}=\R/(2\pi\rho\Z)\times S^1\ ,\quad S^1=
\R/(2\pi \Z)\ .
$$
Again we pass  to the Fourier modes\footnote{Note that, due to the fact that the function $\psi$ is
  real,  $v_\bk=v_{-\bk}^*$.\label{footnote1}} $\{v_\bk,$  $\bk=(m,n)\in
\Z^2\}$
and to the slow time $\tau$ to re-write the equation as 
\begin{equation}\label{eq:CHM}
\begin{split}
\dot v_\bk=-i\nu^{-1}\omega^C_\bk &v_\bk
+\frac{1}{\rho(m^2+n^2\rho^2+F\rho^2)}\sum_{\bk_1,\bk_2\in \Z^2}
\left(m_1^2+n_1^2\rho^2\right)\\ 
&\times\left(m_1n_2-n_1m_2\right)
v_{\bk_1}
v_{\bk_2} \delta^{12}_{\bk} -\gamma_\bk v_\bk
+\mu b_\bk \dot \bb_\bk\ ,
\end{split}
\end{equation}
where $\bk_1=(m_1,n_1)$, $\bk_2=(m_2,n_2)$ and the dispersion function
has the form
\begin{equation}\label{eq:dispCHM}
\omega^C_\bk=-\frac{m\rho}{m^2+n^2\rho^2+F\rho^2}\ ,\quad \bk=(m,n)\in \Z^2\ .
\end{equation}

The effective equations  for eq.~\eqref{eq:NLS} and
eq.~\eqref{eq:CHM} can be  easily obtained on account of the general
formula \eqref{eq:ee}. Using it,  for the NLS equation we get the
effective equation
\begin{equation}\label{eq:eeNLS}
\dot {\tilde a}_\bk=i\gi\sum_{\bk_1,\bk_2,\bk_3\in \Z^d}
\tilde a_{\bk_1} \tilde a_{\bk_2}\tilde a^*_{\bk_3}
\delta^{12}_{3\bk}\delta({\omega^N}^{12}_{3\bk}) -\gamma_\bk \tilde a_\bk
+\mu b_\bk \dot \bb_\bk\,,\quad \bk\in\Z^d\,,
\end{equation}
while for  CHM  the  effective equation is the system 
\begin{equation}\label{eq:eeCHM}
\begin{split}
\dot {\tilde a}_\bk=\frac{1}{\rho(m^2+n^2\rho^2+F\rho^2)}&\sum_{\bk_1,\bk_2\in \Z^2}
\left(m_1^2+n_1^2\rho^2\right)\left(m_1n_2-n_1m_2\right)\\
&\times\tilde a_{\bk_1}
\tilde a_{\bk_2} \delta^{12}_{\bk}\delta({\omega^C}^{12}_\bk)
-\gamma_\bk \tilde a_\bk
+\mu b_\bk \dot \bb_\bk\,,\quad \bk\in\Z^2\,.
\end{split}
\end{equation}

It is clear that the behaviour of solutions for
eqs.~\eqref{eq:eeNLS}-\eqref{eq:eeCHM} is dictated by the structure of
 resonances since  they determine the surviving terms of the
nonlinearity. The geometric properties of the resonant set for the higher dimensional 
NLS  equations are
described in the following section, whereas the resonances
for CHM are discussed in Section~\ref{ssec:resCHM}.

\subsection{Structure of  resonances for the 
NLS equation}\label{ssec:resNLS}
In the case of 2$d$ NLS equation 
 each resonance is formed by four points of
$\Z^2$ which have a simple geometrical characterization: they form the
vertices of a rectangle. Indeed, if a quadruple $\{\bk,\bk_1,\bk_3,\bk_2\}$ 
satisfies \eqref{resonance} with $q=3$, then 
on account of the second relation  we have 
$\bk_1-\bk = \bk_3-\bk_2$. So  the polygonal $\{\bk,\bk_1,\bk_3,\bk_2\}$ is a
parallelogram. Substituting $\bk=\bk_1+\bk_2-\bk_3$ in the first relation 
%resonance condition $\omega^N_{\bk_1}+\omega^N_{\bk_2}=
%\omega^N_{\bk_3}+ \omega^N_\bk$
 and using \eqref{eq:dispNLS} we get 
$$
2(\bk_3\cdot \bk_3+\bk_1\cdot \bk_2-\bk_2\cdot \bk_3-\bk_1\cdot \bk_3)
= 0 \quad \Rightarrow\quad \left(\bk_3-\bk_2\right)\cdot \left(
\bk_3-\bk_1\right) =0\,.
$$
I.e.,  $\bk_3-\bk_2$ is orthogonal to $\bk_3-\bk_1$. So 
$\{\bk,\bk_1,\bk_3,\bk_2\}$ %forms a resonance if and only if
%it constitutes the vertices of 
is a rectangle  in $\Z^2$.

\begin{figure}[h]
\centering
\includegraphics[width=0.5\textwidth]{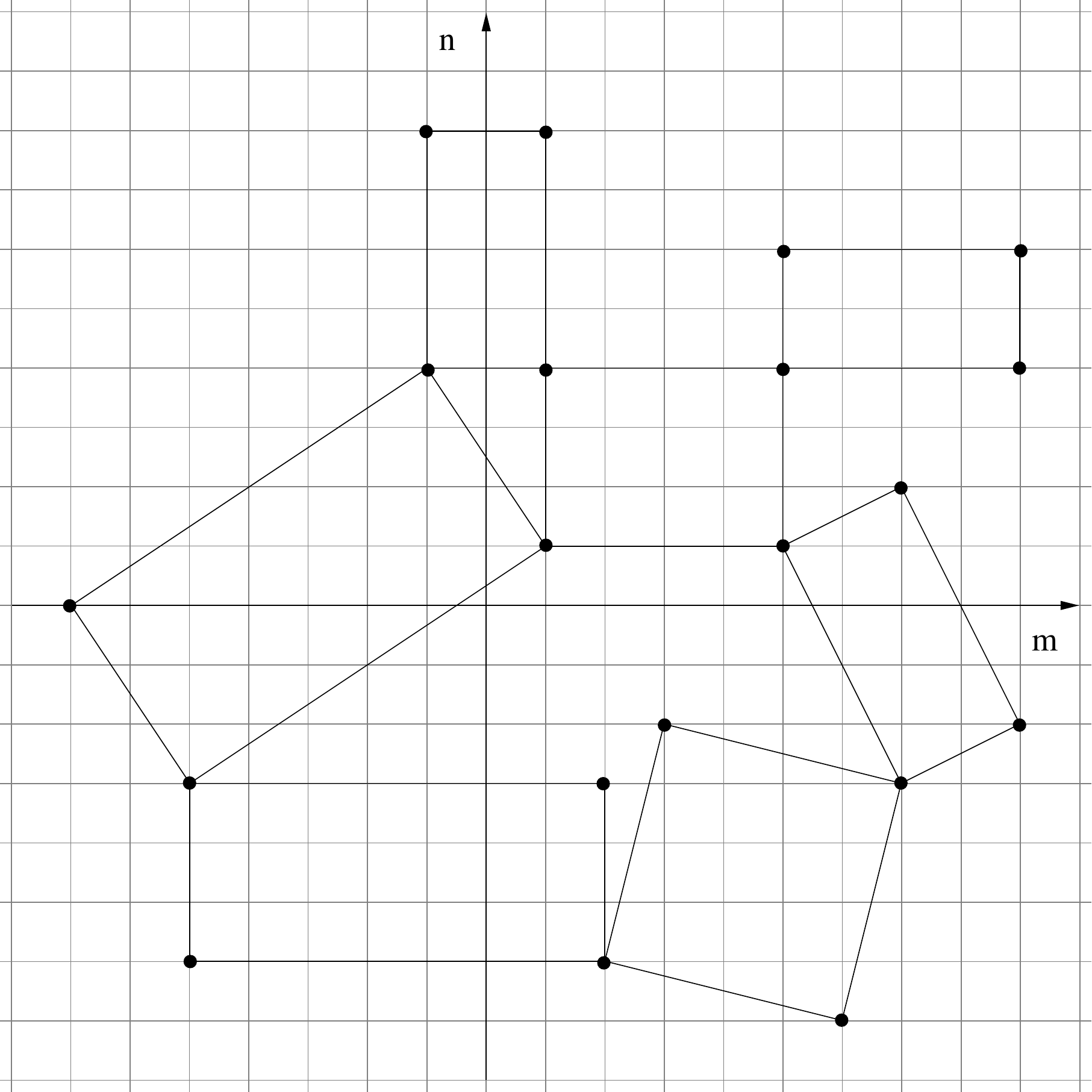}
\caption{Examples of connected resonant quadruples for the NLS
  equation in the $\Z^2$ lattice. Each point is the vertex of at least
one rectangle.}
\label{fig:rect}
\end{figure}
 It is
easy to see that for any vectors $\bk,\bk_1\in\Z^2$ there is an integer rectangle 
of the form $\{\bk, \bk_1, \bk_2, \bk_3\}$.  So the equivalence, defined by the 
clusters of the 2d~NLS equation makes $\Z^2$ a single cluster, and  the 
equation falls in the case 2). A graphical illustration of some
resonant quadruples and their connections in $\Z^2$ is displayed in fig.~\ref{fig:rect}. 

Similar 
all higher-dimensional NLS equations fall in  case 2).

\subsection{Structure of  resonances for CHM}\label{ssec:resCHM}

The structure of resonances for the CHM equation depends on the shape-factor 
$\rho$. Below we discuss it, supposing for  simplicity that the Froude number 
 $F$ is kept fixed (see  \cite{bplane} for the general case).
We start with  explicitly rewriting   for the
present case the resonance condition \eqref{resonance}
 (recall  that $\bk_1=(m_1,n_1)$,
$\bk_2=(m_2,n_2)$ and $\bk=(m,n)$): 
\begin{equation}\label{eq:res_CHM}
\begin{split}
&m_1+m_2=m\ ,\qquad n_1+n_2= n\ ,\\
&\frac{m_1}{m_1^2+n_1^2\rho^2+F\rho^2}+\frac{m_2}{m_2^2+n_2^2\rho^2+F\rho^2}
= \frac{m}{m^2+n^2\rho^2+F\rho^2}\ .
\end{split}
\end{equation}
Solutions $(\bk_1,\bk_2,\bk)$ to these  equations can be
divided to different classes, according to how many numbers
among $m_1,m_2$
and $m$ are non-zero:
\begin{description}
\item[(i)] If all three are zero, then any triad $\bk_1=(0,n_1)$,
  $\bk_2=(0,n_2)$, $\bk=(0,n_1+n_2)$ constitutes a solution. 
  As $c_{\bk_1,\bk_2,\bk}$ vanish in this case (see
  \eqref{eq:eeCHM}), such triads do not form a resonance. 
\item[(ii)] If only one number is different from zero, then \eqref{eq:res_CHM} 
  admits no solution.
\item[(iii)] If only one among $m_1,m_2$ and $m_3$ vanishes, two
  subcases arise (as $\bk_1$ and $\bk_2$ play an exchangeable role):
\begin{description}
\item[a.] if $m_1=0$ (which implies $m_2=m$), then $n_2^2=n^2$ and there are  two
  solutions, one for $n_1=0$, $n_2=n$, and another 
   for
  $n_2=-n_1/2=-n$;
\item[b.] if $m=0$ (which implies $m_1=-m_2$), then $n_1^2=n_2^2$ and again there are  two
  solutions, one for $n=0$, $n_1=-n_2$, and another 
  for $n_1=n_2=n/2$.
\end{description}
\item[(iv)] All three are different from zero. This is the only case
  when the solutions depend on $\rho$. Indeed, let us fix a triad
  $(\bk_1,\bk_2,\bk)$ and look for which values of $\rho$ it
  constitutes a resonance. %By expanding 
  The second line of
  \eqref{eq:res_CHM} may be re-written as
  %we see that it is equivalent to
\begin{equation}\label{eq:poly}
a_0(\bk_1,\bk_2,\bk)+a_1(\bk_1,\bk_2,\bk,F)\rho^2+
a_2(\bk_1,\bk_2,\bk,F)\rho^4 = 0\,,
\end{equation}
where $a_0, a_1$ and $a_2$ are polynomials. In particular, 
$a_0=m_1m_2m\left(m_2 m+m_1m-m_1 m_2\right)$. In  view of the  inequality
$(x^2+y^2+xy)>0$, valid for nonvanishing    $x$ and $y$,
$$
a_0%m_1m_2m\left(m_2 m+m_1m-m_1 m_2\right)
= m_1m_2 m\left( m_1^2+m_2^2+m_1m_2\right)\neq 0\ ,
$$
where the use is made of the first line of \eqref{eq:res_CHM}. This shows
that the second order polynomial in $\rho^2$ at the l.h.s. of
\eqref{eq:poly} is nontrivial. So for any fixed triad $(\bk_1,\bk_2,\bk)$,
where  $m_1$, $m_2$ and $m$ are nonzero,
 relation \eqref{eq:poly} holds 
 for at most two nonnegative values of $\rho$. %Conversely, each triad
%$(\bk_1,\bk_2,\bk)$ with $m_1$, $m_2$ and $m$ all nonzero is a
%solution for \eqref{eq:res_CHM} at most for two nonnegative values of $\rho$.
\end{description}

\begin{figure}[h]
\centering
\includegraphics[width=0.7\textwidth]{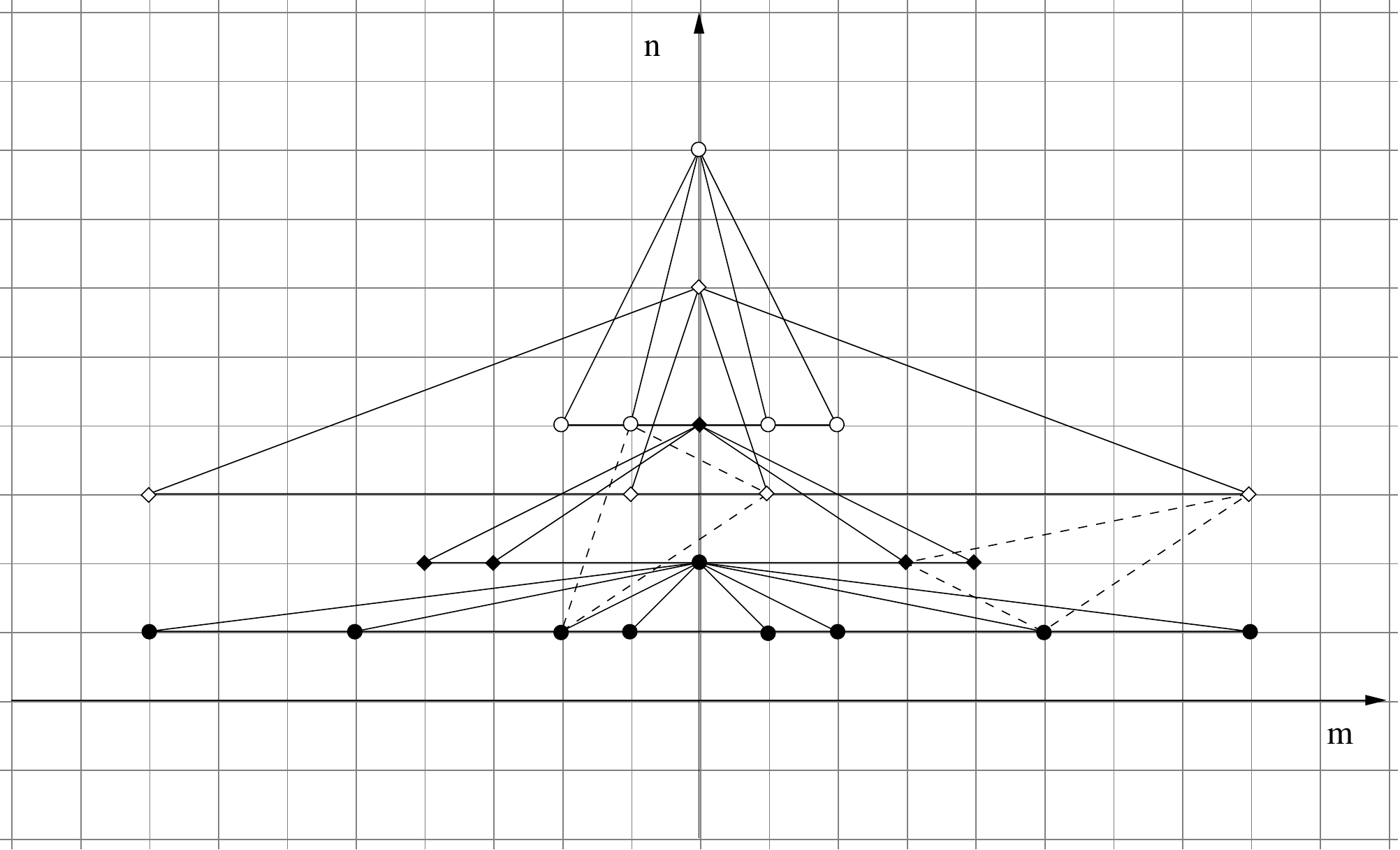}
\caption{Example of connected resonant triads for the CHM
  equation in the $\Z^2$ lattice. Points belonging to different
  clusters of standard 
  resonances are marked with different symbols, solid lines connect
  standard resonances, dashed lines non-standard ones.}
\label{fig:trian}
\end{figure}

The different types of resonances are represented in
Fig.~\ref{fig:trian}, where only the points above the horizontal axis
are displayed (cf. footnote~\ref{footnote1}). There the resonances
of type (iii) (which we will call standard
resonances) are connected by solid lines: they form triangles symmetric
with respect to the vertical axis $m=0$, in which each point $(2m,0)$
is connected with $(m,n)$ and $(m,-n)$, for any $n$. The resonances
of type  (iv) (which we will call non-standard) are
displayed as dashed lines: they constitute triangles in which none of the
vertices lies on the vertical axis.

Since each non-standard resonance appears only for at most two values
of $\rho$, then by removing (at most) a countable set of $\rho$'s we kill all 
 of them. Let us denote 
this removed  set   $\mathcal Z\subset\R_+$. The set 
$\R^+\backslash \mathcal Z$ of
remaining values of $\rho$,  for which no
non-standard resonance appear, can  be regarded as ``typical''. Accordingly we will refer
to $\rho\in \R^+\backslash \mathcal Z$ as \emph{typical} values of $\rho$ (or
as to the {\it typical case}). Below in Section~\ref{sec:CHM} we show that in the 
typical case all  resonances are
divided to non-intersecting clusters of size at most 3, thus fitting the third option, considered in
 Section~\ref{sec:res}.

\section{NLS: the  power-law energy spectrum}\label{sec:NLS}
Effective equation \eqref{eq:eeNLS} for NLS (which, as we have seen,
determines the energy spectrum) is not easy to handle since  its completely  connected
resonance structure (see Section~\ref{ssec:resNLS}) makes
impossible to split it to 
%the system of equations into 
 simpler subsystems (on
 contrary to the  CHM equation,  see Section~\ref{sec:CHM}). We present
here a way to investigate the behaviour of 
solutions of \eqref{eq:eeNLS} when the size $L$ of the box goes to
infinity, based on certain traditional for the wave turbulence heuristic
approximation (see \cite{Naz, ZLF92, New}), following our work
\cite{KM13KZ}. 
This will lead us to a wave kinetic (WK) equation of the
form, usually encountered in the wave turbulence. The treatment follows
closely the paper \cite{KM13KZ}, to which the reader can refer for
further details.

\subsection{The limit  $L\to \infty$}\label{ssec:NLSL}
From the point of view of mathematics, the limit when the size $L$
of the torus $\T^d_L$ tends to infinity in equation \eqref{eq:eeNLS}
presents a serious problem. In particular, for what concerns the
definition of  a possible limiting stochastic equation. Instead of trying to
resolve this difficulty, 
 for any finite $L$ we will study   the expectations  ${\mathbf E}(\tilde I_\bk)$ 
 of the actions $\tilde I_\bk=\tfrac12|\tilde
a_\bk|^2$ of solutions for the corresponding 
equation (the function $\bk\mapsto{\mathbf E}(\tilde I_\bk)$ is called 
called {\it the wave-action spectrum}),  and  then pass to
the limit as $L\to\infty$ only for these quantities. \footnote{In the case of the non-forced 
equation  the expectations should be taken  with respect to the 
distribution of the initial data, while for the forced equation -- with 
respect to the distribution  of the forcing (and, possibly, of the
initial data).}

We fix $L$ and, by making use of Ito's formula for $\tilde I_\bk$,  get from
\eqref{eq:eeNLS} that
\begin{equation}\label{eq:acNLS}
\begin{split}
\frac{d}{d\tau} \tilde I_\bk= &\frac {i\delta}2 \sum_{\bk_1,\bk_2,\bk_3\in
  \Z^d}\left(\tilde a_{\bk_1}\tilde a_{\bk_2}\tilde a^*_{\bk_3}\tilde
a^*_\bk - \tilde a^*_{\bk_1}\tilde a^*_{\bk_2}\tilde a_{\bk_3}\tilde
a_\bk  \right)\delta^{12}_{3\bk}
\delta\left({\omega^N}^{12}_{3\bk}\right) \\
&-\gamma_\bk \tilde
I_\bk+\frac{\mu}2 b_\bk\left( \tilde a^*_\bk\dot \bb_\bk+\tilde a_\bk \dot
\bb^*_\bk\right) + b_\bk^2\ , \quad \bk\in \Z^2\,.
\end{split}
\end{equation}
Now we pass to the expected  values, and define the moment $M^{\bk_1\ldots
  \bk_{n_1}}_{\bk_{n_1+1}\ldots \bk_{n_1+n_2}}(\tau)$ of  $\tilde
a(\tau)$ of order $n_1+n_2$  as
\begin{equation}\label{moments}
M^{\bk_1\ldots \bk_{n_1}}_{\bk_{n_1+1}\ldots
  \bk_{n_1+n_2}}(\tau)=\left\langle
    \tilde a_{\bk_1} \cdots \tilde a_{\bk_{n_1}} \tilde a^*_{\bk_{n_1+1}}\cdots
     \tilde  a^*_{\bk_{n_1+n_2}}\right\rangle_\tau \ ,
\end{equation}
where $\langle \cdot\rangle_\tau$ stands for  the expected value at time
$\tau$, i.e.,
$\langle f(\tilde a)\rangle_\tau={\mathbf E}\big( f(\tilde a(\tau))\big)$ for
any measurable
function $f(\tilde a)$.  Then from the system \eqref{eq:acNLS} we get
\begin{equation}\label{eq:chain_2}
\dot M^\bk_\bk=-2\lla_\bk M_\bk^\bk + 2 b^2_\bk +2\gi \sum_{\bk_1,\bk_2,\bk_3} \Im M^{\bk_1 \bk_2}_{\bk \bk_3}
\delta^{\bk_1\bk_2}_{ \bk \bk_3} \delta({\omega^N}^{\bk_1\bk_2}_{ \bk
  \bk_3})\ ,\quad\bk\in \Z^d\ .
\end{equation}
This system is not closed since it  involves the moments of order
4. By applying again Ito's formula, we can express
the time derivative of  moments of any 
order $n_1+n_2$ as a function of
the moments of order $n_1+n_2-2$ and those  of order $n_1+n_2+2$. The
coupled system, containing the equations for all moments, is called the {\it
chain of moments equation}  (see \cite{MY}).\footnote{Notice that, 
since the equation which we consider has a cubic nonlinearity, 
 equations for  moments of even order are decoupled from those for moments of odd
  order.} 
  Systems of this kind are usually treated  by
approximating  moments of high order by suitable functions of lower
order moments in order to get a closed system of equations. We will
show that if the quasi-stationary and quasi-Gaussian approximations (see below)
are chosen to close the system of moment equations,
then under the limit $L\to \infty$  we recover  a modified version of
the WK equation.

To study the sum in the r.h.s. of \eqref{eq:chain_2}, we notice that
if  the Kr\"onecker deltas
are different from zero  because $\bk$  equals to one vector  among
$\bk_1,\bk_2$ and $\bk_3$ is equal to
another, then the moment is real and does not
contribute to the sum. So we may assume that
 $\bk\neq \bk_1,\bk_2$, $\bk_3\neq \bk_1,\bk_2$. In this case we 
calculate   the fourth order moments in the r.h.s. of
\eqref{eq:chain_2} through Ito's formula (see \cite{KM13KZ}) and get
\begin{equation}\label{eq:chain_4}
\begin{split}
\dot M^{\bk_1 \bk_2}_{\bk \bk_3}&= -(\lla_\bk+\lla_{\bk_1}+\lla_{\bk_2}+\lla_{\bk_3})
M^{\bk_1 \bk_2}_{\bk \bk_3}
+i\gi\sum_{\bk_4,\bk_5,\bk_6} \Bigl(M^{\bk_1 \bk_2 \bk_4}_{\bk_3 \bk_5 \bk_6}
\delta^{\bk \bk_4}_{ \bk_5 \bk_6} \delta({\omega^N}^{\bk\bk_4}_{ \bk_5\bk_6})+ \\
&\!\!\!\!\!\!\!M^{\bk_1 \bk_2 \bk_4}_{\bk \bk_5 \bk_6}
\delta^{\bk_3 \bk_4}_{ \bk_5 \bk_6} \delta({\omega^N}^{\bk_3\bk_4}_{ \bk_5\bk_6})
 - M^{\bk_2 \bk_5 \bk_6}_{\bk \bk_3 \bk_4}
\delta_{\bk_1 \bk_4}^{ \bk_5 \bk_6} \delta({\omega^N}_{\bk_1\bk_4}^{ \bk_5\bk_6}) -  M^{\bk_1 \bk_5
\bk_6}_{\bk \bk_3 \bk_4}
\delta_{\bk_2 \bk_4}^{ \bk_5 \bk_6} \delta({\omega^N}_{\bk_2\bk_4}^{ \bk_5\bk_6} )\Bigr)  \ .
\end{split}
\end{equation}

We make now the first approximation by neglecting the term containing
the time derivative at the l.h.s. of \eqref{eq:chain_4}. This can be
justified, if $\tau$ is large enough, by the quasi-stationary
approximation (cf.  Section~2.1.3 in \cite{ZLF92}).  Namely, let us write equation
\eqref{eq:chain_4}  as
$$
\left(\frac{d}{d\tau}+(\lla_\bk+\lla_{\bk_1}+\lla_{\bk_2}+\lla_{\bk_3})
 \right) M^{\bk_1 \bk_2}_{\bk \bk_3}=f\ .
$$
Notice that since all $\lla_\bk$'s are positive, then the linear differential equation
in the l.h.s. is exponentially stable. Assume that $f$ as a function of $\tau$ is
almost constant during time-intervals, sufficient for  relaxation of the differential
equation. Then
$$
M^{\bk_1 \bk_2}_{\bk
  \bk_3}\approx\frac{f}{\lla_\bk+\lla_{\bk_1}+\lla_{\bk_2}+
  \lla_{\bk_3}} \ .
$$
We insert this in \eqref{eq:chain_2} and get
\begin{equation}\label{eq}
\begin{split}
\dot M^\bk_\bk\approx &-2\lla_\bk M_\bk^\bk + 2 b^2_\bk +2\gi^2 \sum_{\bk_1,\bk_2,\bk_3}
\frac1{\lla_\bk+\lla_{\bk_1}+\lla_{\bk_2}+\lla_{\bk_3}}
\delta^{\bk_1\bk_2}_{ \bk \bk_3} \delta({\omega^N}^{\bk_1\bk_2}_{ \bk \bk_3})\\
&\times\Re\left(\sum_{\bk_4,\bk_5,\bk_6} \Bigl(M^{\bk_1 \bk_2 \bk_4}_{\bk_3 \bk_5 \bk_6}
\delta^{\bk \bk_4}_{ \bk_5 \bk_6} \delta({\omega^N}^{\bk\bk_4}_{ \bk_5\bk_6})+
M^{\bk_1 \bk_2 \bk_4}_{\bk \bk_5 \bk_6}
\delta^{\bk_3 \bk_4}_{ \bk_5 \bk_6} \delta({\omega^N}^{\bk_3\bk_4}_{ \bk_5\bk_6})\right.
 \\&\left.\phantom{\sum_{\bk_4,\bk_5,\bk_6}}- M^{\bk_2 \bk_5 \bk_6}_{\bk \bk_3 \bk_4}
\delta_{\bk_1 \bk_4}^{ \bk_5 \bk_6} \delta({\omega^N}_{\bk_1\bk_4}^{ \bk_5\bk_6}) -
 M^{\bk_1 \bk_5
\bk_6}_{\bk \bk_3 \bk_4}
\delta_{\bk_2 \bk_4}^{ \bk_5 \bk_6} \delta({\omega^N}_{\bk_2\bk_4}^{ \bk_5\bk_6} )\Bigr)\right)
\ .
\end{split}
\end{equation}

We then apply the second approximation, generally accepted in the WT
(see \cite{ZLF92, Fal, Naz, New}) which enables us to transform
the previous relation  to a closed equation for the second order
moments. This consists in the quasi--Gaussian approximation, i.e., in  the
assumption that the higher-order moments \eqref{moments}
can be approximated by polynomials of the second-order moments,
as if the random variables $v_\bk$ were independent complex Gaussian
variables. In particular,
\begin{equation}\label{eq:qgauss}
\begin{split}
M^{{\mathbf{l}}_1{\mathbf{l}}_2{\mathbf{l}}_3}_{{\mathbf{l}}_4
  {\mathbf{l}}_5{\mathbf{l}}_6}\approx 
M_{{\mathbf{l}}_1}^{{\mathbf{l}}_1}
M_{{\mathbf{l}}_2}^{{\mathbf{l}}_2}
M_{{\mathbf{l}}_3}^{{\mathbf{l}}_3}\left(\delta^{{\mathbf{l}}_1}_{{\mathbf{l}}_4}
(\delta^{{\mathbf{l}}_2}_{{\mathbf{l}}_5}\delta^{{\mathbf{l}}_3}_{{\mathbf{l}}_6}
+\delta^{{\mathbf{l}}_2}_{{\mathbf{l}}_6}\delta^{{\mathbf{l}}_3}_{{\mathbf{l}}_5})
\right.
 \left.
+\delta^{{\mathbf{l}}_1}_{{\mathbf{l}}_5}
(\delta^{{\mathbf{l}}_2}_{{\mathbf{l}}_4}\delta^{{\mathbf{l}}_3}_{{\mathbf{l}}_6}
+\delta^{{\mathbf{l}}_2}_{{\mathbf{l}}_6}\delta^{{\mathbf{l}}_3}_{{\mathbf{l}}_4})
+\delta^{{\mathbf{l}}_1}_{{\mathbf{l}}_6}
(\delta^{{\mathbf{l}}_2}_{{\mathbf{l}}_4}\delta^{{\mathbf{l}}_3}_{{\mathbf{l}}_5}
+\delta^{{\mathbf{l}}_2}_{{\mathbf{l}}_5}\delta^{{\mathbf{l}}_3}_{{\mathbf{l}}_4})
\right)\ .
\end{split}
\end{equation}

At this point we pass in equation \eqref{eq}, closed using the
relation \eqref{eq:qgauss}, to the limit $L\to\infty$. This can be
done by approximating sums with integrals if, instead of parametrising 
the modes by integer vectors $\bk\in\Z^d$, we parametrise them by
vectors $\tilde \bk=\bk/L$ from the shrunk   lattice $\Z^d_L=L^{-1} \Z^d$. Accordingly we 
 define  
$$
\tilde M^{\tilde \bk_1\ldots \tilde \bk_{n_1}}_{\tilde
  \bk_{n_1+1}\ldots \tilde \bk_{n_1+n_2}} := M^{\bk_1\ldots
  \bk_{n_1}}_{\bk_{n_1+1}\ldots \bk_{n_1+n_2}}  \ ,\quad \tilde
\gamma_{\tilde \bk}:= \gamma_\bk\ ,\quad \tilde b_{\tilde \bk}:=b_\bk \,,
$$
and note that since the restriction, imposed by the Kr\"onecker deltas, 
 is homogeneous, then it does not change under this re-parametrisation. 
 Abusing notation, we will drop the tildes in the
rest of the Section, but will use the parametrisation by points of  $\Z^d_L$.

We denote by $S_\bk$ the sum, given by 
  the second and third lines of \eqref{eq}, written  in
the new parametrisation, and note that it splits into a
finite number of sums like
$$
S^j_{  \bk}=\sum_{(  \bk_1,  \bk_2,
  \bk_3,  \bk_4,  \bk_5,  \bk_6)\in \Z^{6d}_L\cap
  \Sigma^j_{  \bk}}
F^j_{  \bk}(  \bk_1,  \bk_2,  \bk_3,
\bk_4,  \bk_5,  \bk_6) \,.
$$
Here $\Sigma^j_{  \bk}$ is a manifold in $\R^{6d}$, defined as
\begin{equation*}
\begin{split}
\Sigma^j_{  \bk}=\left\{(\mathbf x_1,\mathbf x_2,\mathbf x_3, \mathbf
x_4,\mathbf x_5,\mathbf x_6)\ : \mathbf x_1+\mathbf x_2= \bk+\mathbf
x_3,
|\mathbf x_1|^2+|\mathbf x_2 |^2=|  \bk|^2+|\mathbf x_3|^2\right.\\
\left.\mathbf x^j+\mathbf x_4=\mathbf x_5+\mathbf x_6,
|\mathbf x^j|^2+|\mathbf x_4 |^2=|\mathbf x_5|^2+|\mathbf x_6|^2,
\mathbf x^j_1=\mathbf x^j_2, \mathbf x^j_3=\mathbf x^j_4, \mathbf
x^j_5=\mathbf x^j_6 \right\}\ ,
\end{split}
\end{equation*}
where $\mathbf x^j$ stands for one among the vectors $  \bk, \mathbf x_1,\mathbf
x_2, \mathbf x_3$, and $\{\mathbf x^j_1,\ldots,\mathbf x^j_6\}$ -- for a
permutation of the set $\{  \bk, \mathbf x_1,\ldots,\mathbf
x_6\}\backslash\{\mathbf x^j\}$.\footnote{Note that the relations, defining
  $\Sigma_{  \bk}^j$ are
not independent.}
 It
is easy to see that since every   $F^j$ is a regular function, then
when  passing from the sums to integrals in the limit  $L\to \infty$,  each term
$S^j_{  \bk}$ as a function of $L$ becomes proportional to  $L^m$, where $m$ is the dimension of the
manifold $\Sigma^j_{ \bk}$. A detailed analysis of all cases shows that the
terms of the highest order in $L$ in the integral
correspond to terms
of the form
$$
S^j_{  \bk}=\sum_{  \bk_1,  \bk_2,  \bk_3}
F_{  \bk}(  \bk_1,  \bk_2,  \bk_3) \delta^{
  \bk_1  \bk_2}_{  \bk   \bk_3}
\delta({\omega^N}^{  \bk_1   \bk_2}_{  \bk   \bk_3})
$$
in the sum $S_{\bk}$, where $\vec k:= (\bk_1,\bk_2,\bk_3)\in
\Z^{3d}_L=:\cM$. Denote
$$
\Sigma_{\bk}=\left\{\vec x= (\mathbf x_1,\mathbf x_2,\mathbf x_3)\in
\R^{3d}: \mathbf x_1+\mathbf x_2=\bk+\mathbf x_3,
|\mathbf x_1|^2+|\mathbf x_2 |^2=|\bk|^2+|\mathbf x_3|^2\right\}\ .
$$
This is a manifold of dimension $3d-d-1=2d-1$, smooth outside the origin. The
latter lies  outside $\Sigma_{\bk}$ if $\bk\neq 0$, and is a singular point
of $\Sigma_{\bk}$ if $\bk=0$.

As shown in \cite{KM13KZ}, in the limit  $L\to \infty$ the sum 
$S_\bk$ can be approximated  by the integral
\begin{equation}\label{sumint}
S_{\bk}\approx L^{2d-1} \int_{\Sigma\backslash \{0\}}
\frac{F_{\bk}(\vec
  x)}{\vp_{\bk}(\vec x)} d \vec x\ ,
\end{equation}
where $\vp_{\bk}(\vec x)$ is a certain  function  on $\Sigma_{\bk}$,
smooth outside zero, such that
\begin{equation}\label{eq:propphi}
\begin{split}
&V_1\le \vp_{\bk}(\vec x)\le V_1(3d)^{d-1/2}\ ,\quad  \vp_\bk(\vec
x):=\vp_{m\bk}(m\vec x)\ ,\\& \vp_\bk(\mathbf x_1,\mathbf x_2,\mathbf x_3)
= \vp_{\bk}(\mathbf x_2,\mathbf x_1,\mathbf x_3)\ ,\quad
\vp_\bk(\mathbf x_1,\mathbf x_2,\mathbf x_3)
= \vp_{\mathbf x_3}(\mathbf x_1,\mathbf x_2,\bk)\,,
\end{split}
\end{equation}
where $V_1$ is the volume of the $1$-ball in $\R^{2d -1}$.

By substituting \eqref{eq:qgauss} in \eqref{eq} and using
\eqref{sumint} we get the limiting (as $L\to\infty$)
 equation in the form
\begin{equation*}
\begin{split}
\dot M^\bk_\bk\approx &-2\lla_\bk M_\bk^\bk + 2 b^2_\bk +4\gi^2L^{2d-1}
\int_{\R^{3d}\backslash\{0\}} d\bk_1 d \bk_2 d
\bk_3  \frac{\vp_\bk^{-1}(\bk_1,\bk_2,\bk_3)}{\lla_\bk+\lla_{\bk_1}+\lla_{\bk_2}+\lla_{\bk_3}}
\delta^{\bk_1\bk_2}_{ \bk \bk_3} \\
&\times\delta({\omega^N}^{\bk_1\bk_2}_{ \bk \bk_3})
\Bigl( M_{\bk_1}^{\bk_1} M_{\bk_2}^{\bk_2} M_{\bk_3}^{\bk_3}+ M_{\bk}^{\bk}
M_{\bk_1}^{\bk_1} M_{\bk_2}^{\bk_2}
- M_{\bk}^{\bk} M_{\bk_2}^{\bk_2}
M_{\bk_3}^{\bk_3}- M_{\bk}^{\bk} M_{\bk_1}^{\bk_1} M_{\bk_3}^{\bk_3}\Bigr)\ .
\end{split}
\end{equation*}

Finally, we define
\begin{equation}\label{eq:scal_n_b}
n_\bk=L^d M_\bk^\bk/2\ , \quad \mathfrak b_\bk=L^{d/2} b_\bk\ ,
\end{equation}
 (so that $\sum_\bk
M_\bk^\bk/2\to \int n_\bk$ and $\sum_\bk b^2_\bk\to \int \mathfrak b^2_\bk$
as $L$ goes to infinity), choose
\begin{equation}\label{eq:scal_rho}
\gi(L)={\tilde \eps}^2L^{1/2}=\frac{\eps^{2q*}}{\nu}{\tilde \eps}^2L^{1/2}\ ,
\end{equation}
for some $\tilde \eps>0$, and get the kinetic equation 
\begin{equation}\label{eq:kz_1}
\begin{split}
\dot n_\bk&= -2\lla_\bk n_\bk+ \mathfrak b^2_\bk + 16{\tilde \eps}^4
\int_{\R^{3d}\backslash\{0\}} d \bk_1\,d \bk_2\, d\bk_3
\delta^{\bk_1\bk_2}_{ \bk \bk_3} \delta({\omega^N}^{\bk_1\bk_2}_{ \bk \bk_3}) 
\\
&\times\frac{\vp_\bk^{-1}(\bk_1,\bk_2,\bk_3)}{\lla_\bk\!+\!\lla_{\bk_1}\!+\!\lla_{\bk_2}\!+\!\lla_{
    \bk_3}} \Bigl(n_{\bk_1}n_{\bk_2}n_{\bk_3}+ n_{\bk}
n_{\bk_1}n_{\bk_2}
- n_{\bk} n_{\bk_2}
n_{\bk_3}- n_{\bk} n_{\bk_1} n_{\bk_3}\Bigr)\,.
\end{split}
\end{equation}
%where the  smooth outside the origin
% function   $\vp_{\bk}(\vec x)$  satisfies  \eqref{eq:propphi}.
 We have thus shown that, with the proper scaling of $\gi$ and $b$ given by
  \eqref{eq:scal_n_b}-\eqref{eq:scal_rho}), the function $n_\bk$ satisfies a
  kinetic equation,   similar to the WK equation for 
NLS  in the classical wave turbulence theory
 (see, for instance, formula (6.81) of \cite{Naz}, where
$d=2$). The differences
 are two: obviously in our case there are  forcing and 
 dissipation,  absent in  the traditional WK
 equations. More interesting is  the  nonvanishing denominator
$\lla_\bk+\lla_{\bk_1}+\lla_{\bk_2}+\lla_{\bk_3}$ which regularises  the integral
 since it growths to infinity with $\bk$, 
and which  modifies the spectra.

\subsection{Power law spectra}\label{ssec:NLSKZ}
Now, under some additional 
 approximation and using the well known Zakharov argument (see
\cite{ZLF92,Naz,New})  we will  get stationary solutions of  eq.~\eqref{eq:kz_1}
with power law energy spectra (more
properly,  wave-action spectra) $\{n_\bk\}$.

To do this  we have to restrain our analysis  to the inertial
interval, i.e., to the spectral
interval, where the  damping and   forcing are negligible.
That is, we should consider eq.~\eqref{eq:kz_1}, supposing that the 
 wave-vector $\bk$ belongs to a sufficiently large spectral region, where 
the first two terms at the r.h.s. of \eqref{eq:kz_1} can be
neglected,  compared  with the third. This happens, e.g., if there the 
solution $\{n_\bk\}$ is 
of order one, while $\mathfrak b_\bk\ll 1$ and
$\gamma_\bk \ll 1$ (i.e., the damping and the dissipation are small at
that spectral region). In  the inertial interval we end up with the
equation
\begin{equation}\label{eq:kz_2}
\begin{split}
\dot n_\bk\approx &\,16\, {\tilde \eps}^4
\int_{\R^{3d}\backslash\{0\}} d \bk_1\,d \bk_2\, d\bk_3
\delta^{\bk_1\bk_2}_{ \bk \bk_3} \delta({\omega^N}^{\bk_1\bk_2}_{ \bk \bk_3})
\frac{\vp_\bk^{-1}(\bk_1,\bk_2,\bk_3)}{\lla_\bk+\lla_{\bk_1}+\lla_{\bk_2}+\lla_{
    \bk_3}}
\\
&\times \Bigl(n_{\bk_1}n_{\bk_2}n_{\bk_3}+ n_{\bk}
n_{\bk_1}n_{\bk_2}
- n_{\bk} n_{\bk_2}
n_{\bk_3}- n_{\bk} n_{\bk_1} n_{\bk_3}\Bigr)\,.
\end{split}
\end{equation}
Notice that, while in the inertial interval we can simply approximate
$\mathfrak b_\bk$ with zero,
this  cannot be done to $\lla_\bk$ since these numbers 
appear in the
denominator of the integral at the r.h.s. of \eqref{eq:kz_1} (their sum makes  the denominator 
of the 
so-called {\it collision term}), and  play an essential role in determining  of the spectrum.

The previous equation has  the form of the four-wave kinetic
equation (see, for instance,  formula 2.1.29 of \cite{ZLF92}). It is
well known (see \cite{ZLF92,Naz}) how to solve such  equations for
stationary spectra with the aid of the Zakharov transformations, provided that 
 the terms
$$
\cT^{\bk,\bk_3}_{\bk_1,\bk_2} =
\frac{\vp_\bk^{-1}(\bk_1,\bk_2,\bk_3)}{\lla_\bk+\lla_{\bk_1}+\lla_{\bk_2}+\lla_{
    \bk_3}}
$$
satisfy, for some  $m\in \R$, the following  conditions of symmetry and homogeneity:
\begin{eqnarray*}
\cT^{\bk,\bk_3}_{\bk_1,\bk_2}= \cT^{\bk_3,\bk}_{\bk_1,\bk_2}=
\cT^{\bk,\bk_3}_{\bk_2,\bk_1}=\cT^{\bk_1,\bk_2}_{\bk,\bk_3}\ ,\qquad 
\cT^{\lambda \bk,\lambda\bk_3}_{\lambda\bk_1,\lambda\bk_2}= \lambda^m
\cT^{\bk,\bk_3}_{\bk_1,\bk_2} \,.
\end{eqnarray*}

Since $\vp$ is a homogeneous function of degree 0 due to
\eqref{eq:propphi}, the requirements above are met if
 on the inertial range the function 
  $\lla_\bk$ can be approximated by a  homogeneous function of the form
$\lla_\bk=\eps' |\bk|^m$, where $m$ is a real number and 
$\eps'\ll 1$ is a small parameter to
guaranty  that the dissipation term  indeed is negligible.

We abbreviate $|\bk|=k$ and 
 look for stationary isotropic, spectra behaving as  power laws of $k$, 
i.e.  $n_\bk=n_k\propto k^{\nu}$ for some real $\nu$,
 by searching $\nu$ such that the r.h.s. of \eqref{eq:kz_2}
vanishes.  The result (see \cite{KM13KZ}) is that, in addition to the
equilibrium solutions $n_k=C$  and $n_k=C/k^2$, which correspond,
respectively, to the equipartition of the wave action and of the
quadratic energy (Rayleigh-Jeans distribution), two nontrivial
power law stationary distributions appear. These are
the  solutions:
\begin{equation}\label{exponents}
n_k\propto k^{-(m+3d-2)/3}\ ,\qquad  n_k\propto k^{-(m+3d)/3}\ .
\end{equation}
If $m=0$, they 
 coincide with the well known in the wave turbulence power-law spectra for 
 the free NLS equation \eqref{eq:ZL*}${}_{\nu=0}$
 (for $d\ge 2$, see \cite{Naz}),
 but the dissipation modifies the power law of the decay if $m\neq 0$.

\section{CHM: resonance clustering}\label{sec:CHM}
Let us  consider in more detail the effective equation
\eqref{eq:eeCHM} for the CHM equation for typical values  of
the shape-factor  $\rho$. 
By the definition of a typical $\rho$  (see
Section~\ref{ssec:resCHM}), no resonances corresponding to
the case (iv) of Section~\ref{ssec:resCHM} occur.
% Moreover, it is
%easy to see that the resonances of the type (i) give no contribution to
%the nonlinearity $R_\bk$ at the r.h.s. of \eqref{eq:eeCHM}, since the
%corresponding terms vanish.
We can then write
the effective equation explicitly, following \cite{bplane}. It will
only involve resonances of type (iii).

Let us consider the equations for the variables  $\tilde a_\bk$ with
$\bk=(m,n)$, separating the cases  $m=0$ and $m\ne0$. When  $m=0$,
the only terms which survive in the nonlinearity $R_\bk(\tilde a)$  are those
where  $\bk_1$ and $\bk_2$ satisfy the  relation (iii-b) of
Section~\ref{ssec:resCHM}, while for $m\neq 0$ only the terms falling in the
case (iii-a) give contribution. For $m=0$, the
nonlinearity vanishes  if $n$ is odd, while  if it is even, then
$$
R_\bk(\tilde a)= \frac 1 {\rho(m^2+n^2\rho^2+F\rho^2)} \sum_{m_1\in \Z}
\left(m_1^2 + \frac{n^2\rho^2}4\right)  m_1 n
\,\tilde a_{(m_1,n/2)}\,\tilde a_{(-m_1,n/2)}\ ,\quad m=0\ ,
$$
which in turn vanishes because  the odd symmetry in $m_1$. On the other
hand, if $m_1\ne0$, then  $\bk_1$ and $\bk_2$ are completely determined
by $\bk$. So we  get that
 \begin{equation}\label{explicit}
R_\bk(\tilde a)=  \left(\frac{2mn}{\rho(m^2+
  n^2\rho^2+F\rho^2)}\left(3n^2\rho^2-m^2\right) \tilde a_{\bar\bk}
\,\tilde a_{(0,2n)} \right)\,,
\end{equation}
where we denoted  $ \bar\bk:= (m,-n)$. Note that this formula applies 
 for the both case $m=0$ and  $m\ne0$.

Expression \eqref{explicit} entails the remarkable consequence that
the hamiltonian part of the  effective equation, i.e., the system in
which forcing and dissipation are removed,
\begin{equation}\label{explicit_nonstoch}
\frac{d}{d\tau} \tilde a_\bk=R_\bk(\tilde a)\ , \quad \bk\in \Z^2\ ,
\end{equation}
is integrable and decomposes to invariant subsystems of complex dimension
at most three.
Indeed, if    $m$ or $n$ vanish, then $R_\bk=0$
and $\tilde a_\bk(t)=\const$. Now let $m, n\ne0$. If $3\rho^2n^2=m^2$, then
again the equation for $\tilde a_\bk$ trivialises. Suppose that $3L^2n^2\ne
m^2$ and denote
\be\label{Ak}
A_\bk= \frac{2mn}{m^2+ n^2\rho^2+ F\rho^2}\left(3n^2\rho^2-m^2\right) \in \R\,.
\ee
Then $A_{\bar\bk}\equiv - A_\bk $. Eq.~\eqref{explicit_nonstoch} (with
any fixed $\bk$)  belongs to the following invariant  sub-system of
\eqref{explicit_nonstoch}:
 \be \label{syst}
 \begin{split}
 &\frac{d}{d\tau}\tilde a_\bk=A_\bk\, \tilde a_{(0,2n)} \tilde a_{\bar\bk},  \\
 &\frac{d}{d\tau} \tilde a_{\bar\bk} = -A_\bk\, \tilde a^*_{(0,2n)} \tilde
   a_\bk ,\\
 &\frac{d}{d\tau} \tilde a_{(0,2n)} =0
 \end{split}
\ee
(we recall that $\tilde a_{(0,2n)}=\tilde a^*_{(0, -2n)}$ by the reality
condition, see footnote \ref{footnote1}).  This
system is explicitly soluble:  if $\tilde a_{(0,2n)}(0)\ne0$, then
\be\label{vk}
\begin{split}
&\tilde a_{(0,2n)}(t)=\,\mathop{\rm Const}\,,\\
&\tilde a_\bk(t)=  \tilde a_\bk(0) \cos(|A_\bk \tilde a_{(0,2n)}| t)
  +\tilde a_{\bar\bk}(0) {\,\sgn}
(A_\bk \tilde a_{(0,2n)})  \sin(|A_\bk \tilde a_{(0,2n)}| t)\,,
\end{split}
\ee
 where for a complex number $z$ we denote
$$
\sgn (z) = z/|z|\; \text{if $z\ne0$, and } \  \sgn (0)=0\,.
$$
The formula for $\tilde a_{\bar\bk}(t)$ is obtained from that for
$\tilde a_\bk(t)$
by  swapping $\bk$ with  $\bar\bk$ and replacing $\tilde a_{(0,2n)}$ by its
complex conjugate. All these solutions are periodic, and it is easy to check 
 that $|\tilde a_\bk|^2+|\tilde a_{\bar \bk}|^2$ and $|\tilde
a_{(0,2n)}|^2$ are integrals  of motion for eq.~\eqref{syst}.

We have  established that there is no Hamiltonian exchange of
energy between  different modes, apart the coupled modes $\tilde a_\bk$ and
$\tilde a_{\bar \bk}$. The situation does not change much when we switch in the 
forcing and the dissipation since  the effective
equation \eqref{eq:eeCHM}, too, splits  to invariant
 subsystems of complex dimension one (if $mn=0$ or
 $3n^2\rho^2=m^2$),  or of dimension
three (otherwise). These systems either are independent, or have {\it
  catalytic interaction} through the variables
$\tilde a_{(0,2n)}$, which  satisfy the Ornstein--Uhlenbeck equation
$$
\frac{d}{d\tau} \tilde a_{(0,2n)} = - \gamma_{(0,2n)} v_{(0,2n)} +\mu
b_{(0,2n)}  \dot\bb_{(0,2n)} \,,
$$
and are independent from other variables.

Being particularly interested in the exchange of energy, let us
consider the equation for the actions $\tilde I$. Due to the
conservation of $|\tilde a_\bk|^2+|\tilde a^2_{\bar \bk}|^2$, Ito's
formula gives (cf.~\eqref{eq:acNLS})
$$
\frac{d}{d\tau} \left(\tilde I_\bk+\tilde I_{\bar \bk}\right) =
-\gamma_\bk \tilde I_\bk-\gamma_{\bar \bk} \tilde I_{\bar \bk} +\frac
\mu 2 b_\bk\left(\tilde a^*_\bk \dot\bb_\bk+\tilde a_\bk\dot
\bb_\bk^*\right)  +\frac
\mu 2 b_{\bar \bk}\left(\tilde a^*_{\bar \bk} \dot\bb_{\bar\bk}+\tilde a_{\bar\bk}\dot
\bb_{\bar\bk}^*\right) +b_\bk^2+b_{\bar \bk}^2\ .
$$
By taking the expected  value, we see that  the second order 
moments satisfy
\begin{equation}\label{eq:momCHM}
\begin{split}
\dot M_\bk^\bk+\dot M_{\bar \bk}^{\bar \bk} = -\gamma_\bk M_\bk^\bk
-\gamma_{\bar \bk} M_{\bar \bk}^{\bar \bk} +2(b_\bk^2+b_{\bar
  \bk}^2)\ ,
\dot M^{(0,n)}_{(0,n)}=-\gamma_{(0,n)} M^{(0,n)}_{(0,n)}+b^2_{(0,n)}
\end{split}
\end{equation}
This equations should be compared with \eqref{eq:chain_2}: they show
that the amount of energy contained in a given cluster is not transferred to
other clusters and depends only on the forcing and the dissipation,
acting in its interior. Thus the energy cascades cannot occur for 
\emph{typical} values of $\rho$. 

\section{Discussion}\label{sec:disc}
In this Chapter we presented a method to study a weakly nonlinear PDE
by investigating  properties of the corresponding \emph{effective equation},
written in terms of the nonlinearity and the resonances in the spectrum of 
the linear part of the equation. 

We have considered two examples of  equations, where the structures of resonances
are completely different. Namely, for 
 the NLS equation  all resonances are connected, and for it we 
  have provided a way of getting power law stationary
energy spectra  (evoking a heuristic physical argument
in addition to a rigorous mathematical theory),
while for the Charney--Hasegawa--Mima equation the resonances form finite clusters, and for 
this equation  we have shown (completely rigorously)
 that, in the \emph{typical} case, no exchange of energy between different oscillating modes
 occurs.

\bibliography{meas}
\bibliographystyle{spphys}

\end{document}